\documentstyle[floats,aps,prl,preprint,psfig]{revtex}

\begin{document}
\draft
\title{ Local Isoelectronic Reactivity of Solid Surfaces}

\author{Steffen Wilke, Morrel H. Cohen} 
\address{Corporate Research Laboratories, Exxon Res. \& Eng. Comp., Annandale,
NJ 08801, USA}
\author{and Matthias Scheffler}
\address{Fritz-Haber-Institut der Max-Planck-Gesellschaft,
Faradayweg 4-6, D-14195 Berlin-Dahlem, Germany.}

\date{\today}

\maketitle

\begin{abstract}
The quantity  
$w^{N}(\text{\bf r}) =
( 1 /  k^2 T_{\text{el}})
[ \partial n(\text{\bf r}, T_{\text{el}}) /
\partial T_{\text{el} } ]_{v({\bf r}),N}$
is introduced as a convenient measure of the local isoelectronic reactivity of
surfaces. It characterizes the local polarizability of the surface and it can
be calculated easily.  The quantity $w^{N}(\text{\bf r})$ supplements the
charge transfer reactivity measured e.g. by the local softness to which it is
closely related. We demonstrate the applicability and virtues of the function
$w^{N}(\text{\bf r})$ for the example of hydrogen dissociation and adsorption
on Pd\,(100).
\end{abstract}

Density-functional calculations of chemisorption processes and of potential
energy surfaces of the dissociation of simple molecules over surfaces, which
have become available in the last years, have greatly improved the fundamental
understanding of reactions at solid surfaces. However, those extensive
computations are limited to a restricted number of model systems. It remains
an important task to develop a methodology that allows the prediction and
interpretation of reactions at surfaces in terms of the properties of the
non-interacting systems.  Such concepts, well known in molecular chemistry,
are the essence of ``reactivity theory''. They are based on low-order
perturbation theory and aim at a description of the early stages of chemical
interactions. The ensuing response functions are called ``reactivity indices''
and characterize the changes of the electronic structure of one reactant as
stimulated by the presence of the other or vice versa. In an early
contribution to this field Fukui {\em et al}.~\cite{fuk52,fuk82} established
the correlation between the frontier-orbital density, {\em i.e.}, the density
of the highest occupied and lowest unoccupied molecular orbital (HOMO and
LUMO), and the reactivity of a system towards electron donation or
accptance. Pearson~\cite{pear63} introduced the electronic ``softness'', the
magnitude of the change of the electronic structure due to a change of the
number of electrons in the system, as a measure of the reactivity.  Species
are classified as ``soft'' if only a small energy is required to change their
electronic configuration, {\em i.e.}, if the valence electrons are easily
distorted, polarized, removed or added.  A ``hard'' species has the opposite
properties holding its valence electrons more tightly~\cite{pear63,parr83}.
The utility of the hardness--softness concept is based on the so called hard
and soft acid and base (HSAB) principle formulated by Pearson~\cite{pear63}
which states that hard-hard (soft-soft) interactions are preferred.  In the
case of polyatomic or extended systems the HSAB principle is used in a local
version: The soft (hard) parts of one reactant prefer to interact with soft
(hard) areas of the other~\cite{parr89}.  Parr and collaborators~\cite{parr89}
gave a foundation in density functional theory to those mostly semi-empirical
concepts, and Cohen {\em et al.}~\cite{coh96a,coh94b,coh94c} have reviewed the
foundations of reactivity theory and addressed some unresolved issues.

The (local) softness and
hardness describe the response of the electron density to a change of the
charge state of the system.  For extended, gapless systems the local softness
$s(\text{\bf r})$ is defined as
\begin{eqnarray}
s(\text{\bf r}) &= &\left( {\partial n(\text{\bf r}) \over \partial \mu}
\right)_{v(\text{\bf r}),T_{\rm el}}  
\label{soft}
\end{eqnarray}
where $v(\text{\bf r})$ is the electrostatic potential due to the nuclei, and
${\em T_{\rm el}}$ is a temperature (see below).
The global softness, $S$, is defined as the integral of the local softness
over all space. The local softness is the local density of states at the
Fermi energy, $g(\text{\bf r},\mu)$, screened by a response function closely
related to the static dielectric function~\cite{coh94b}; it describes the {\em
local} response of the electron density to a {\em global} change of the charge
state of the system.  The chemical interaction will be, however, largely
determined by the {\em nonlocal} response of the electron density to a {\em
local} perturbation, and is governed by the whole spectrum and not just the
states at the Fermi energy~\cite{coh96a}.  Despite these restrictions, the
local softness, in connection with the HSAB principle, has become an important
predictive tool in analyzing reactions between molecules~\cite{lee87,lan90}.

>From the beginning, application of the softness concept to surface processes
and, in particular, to catalytic reactions at metal surfaces has been
proposed, and empirical rules, {\em e.g.} for the influence of adsorbate
layers on the surface properties of metals, have been
deduced~\cite{pear63,dron91,yang85,stair91}.  However, for extended systems
molecular orbitals and levels are not well defined and models of chemical
reactions at surfaces are reformulated using a basis of localized atom-like
orbitals and the projected density of states (see
e.g.~\cite{paul60,hof88,coh94a,ham95}).  Only recently have investigations
appeared which apply the HSAB principle to semiconductor
systems~\cite{pin93,gal93,piq94,brom94}.  Calculations of the local softness
at metal surfaces have not yet been done so far.  Metal surfaces are
characterized by a non-zero density of states (DOS) at the Fermi level, and
screen external perturbations efficiently by low-energy electron-hole
excitations.  In the framework of reactivity theory the metal surface is thus
soft. In much the same spirit, Feibelman and Hamann~\cite{feib84a} related the
change of reactivity of transition metal surfaces in the presence of adsorbate
layers to the spatial variation of the DOS at $E_{\text F}$.  Although their
study demonstrated that the states in the vicinity of the Fermi level govern
the reactivity of metal surfaces, no clear correlation between local
reactivity and the spatial variation of the DOS close to $E_{\text F}$ could
be established. Falicov and Somorjai~\cite{fal85} stressed the importance of
low-energy occupation fluctuations for the large reactivity of metal
surfaces. Yang and Parr~\cite{yang85} as well as Baekelandt et al.~\cite{baek95}
have shown that the local softness is directly related to the correlation
function of changes of the total number of electrons and the density
fluctuations.

In this paper we introduce the change of the electron density due to the 
excitation of low-energy electron-hole pairs induced by an increased
electron temperature, $T_{\rm el}$, as a quantity which characterizes the
spatial distribution of the reactivity of metal surfaces. 
We show that this
isoelectronic
reactivity index, denoted as
$w^N(\text{\bf r})$, represents 
a convenient  measure of the capability of the metal
electrons to react in response to incoming atoms or molecules:
\begin{eqnarray}
\label{wdef1}
w^N(\text{\bf r})
& = & { 1 \over  k^2
T_{\text{el}} } \ \left( {\partial n (\text{\bf r},T_{\text{el}}) \over \partial 
T_{\text{el}}} \right)_{v({\bf r}),N} \\
\label{wdef2}
& \approx & { n(\text{\bf r},  T_{\text{el}} ) - n(\text{\bf r}, 0)
\over (k T_{\text{el}})^2 } \quad ;
\end{eqnarray}
where $N$ is the number of electrons in the system.

The local softness and the function $w^N(\text{\bf r})$ are closely related,
as may be concluded already from their definitions in equ.~(\ref{soft}) and
equ.~(\ref{wdef1}).  Below, a formal expression for the function
$w^N(\text{\bf r})$ is presented showing this connection and the physical
significance of $w^N(\text{\bf r})$ more clearly.  We define in addition to
the isoelectronic function $w^N(\text{\bf r})$ a function $w^\mu(\text{\bf
r})$ by taking the derivative in equ.~(\ref{wdef1}) at constant Fermi level
$\mu$ instead of at constant number of electrons. Both functions are related
by
\begin{eqnarray}
w^N(\text{\bf r})
& = & w^\mu(\text{\bf r}) 
- s(\text{\bf r}) { W^\mu \over S }  
\label{mun}
\end{eqnarray}
where $s(\text{\bf r})$ and $S$ are the local and global softnesses and
$W^\mu$ is the integral of $w^\mu(\text{\bf r})$ over
space. Equation~\ref{mun} shows that the integral of $w^N(\text{\bf r})$ over
space is zero, as expected for an isoelectronic reactivity function. The
density, $ n(\text{\bf r}, T_{\rm el})$, entering Eq.~\ref{wdef2} may be expressed as the
integral of the product of the Fermi function, $f(\epsilon)$, and the local
Kohn-Sham density of states $g(\text{\bf r},\epsilon)$ over the energy
$\epsilon$. In this paper we use the Fermi function at a non-zero auxiliary
temperature, $T_{\text{el}}$.  Applying the same technique as has been
described for the local softness in Ref.~\onlinecite{coh94b}, the function
$w^\mu(\text{\bf r})$ is the solution of an integral equation of the same
structure as that derived for the local softness~\cite{coh94b,coh96a}
\begin{eqnarray}
w^\mu(\text{\bf r})
& = & \int d^3 {\bf r'} \ K^{-1}({\bf r},{\bf r'}) \ i({\bf r'})\quad .
\label{inte}
\end{eqnarray}
In the limit of zero electron temperature $T_{\text{el}}$, the kernel in
equ.~(\ref{inte}) is the same response function as found for the local
softness~\cite{coh94b}. The source term $i({\bf r})$
\begin{eqnarray}
i({\bf r}) & = & { 1 \over  k^2 T_{\text{el}} } \int d\epsilon \  \left(
{\partial
 f(\epsilon) \over \partial T_{\text{el}}} \right)_{v({\bf r}),\mu} \
g(\text{\bf r},\epsilon)  \\ 
& \stackrel{T_{\text{el}} \rightarrow 0}{\longrightarrow}
& { \pi^2 \over 3 } \ \left( {\partial  g(\text{\bf r},\epsilon) \over
\partial \epsilon } \right)_{\epsilon=\mu}  
\label{source}
\end{eqnarray}
is proportional to the energy derivative of the local density of states at the
Fermi energy. Thus, the function $w^\mu(\text{\bf r})$ is essentially a
(short-range) linear mapping of the energy derivative of the local DOS at the
Fermi level.  The local softness, on the other hand, is the same mapping but
applied to the local DOS at the Fermi level~\cite{coh94b,coh96a}.

As mentioned in the introduction the local softness measures the correlation
between density fluctuations and fluctuations in the total number of electrons
in a statistical ensemble~\cite{yang85,baek95}. In a grand-canonical ensemble
the reactivity index $w^N(\text{\bf r})$ has an interesting relationship to
the correlation function between the density fluctuations and the fluctuations
of the total energy of the surface:
\begin{eqnarray}
w^N(\text{\bf r})&  = &{ 1 \over ( k T_{\text{el}})^3 } \: \{  \langle \left( H -
\langle H \rangle \right) \left( n(\text{\bf r}) - \langle  n(\text{\bf r})
\rangle \right) \rangle \nonumber \\ 
& - & {s((\text{\bf r}) \over S} \ \langle \left( H -
\langle H \rangle \right) \left( N - \langle  N
\rangle \right) \rangle \}  \quad ,
\end{eqnarray}
where $H$ is the Hamiltonian of the system and $\langle \rangle$ indicates the
averaging over the grand-canonical ensemble.

The quantities $w^\mu(\text{\bf r})$ and $w^N(\text{\bf r})$ describe the
spatial variation of the strength and the sign of the density change connected
with low-energy excitations of the valence electrons.  They represent
essentially the difference of the densities of the LUMOs and HOMOs: Regions of
positive values of $w^N(\text{\bf r})$ correlate to dominant spatial weight of
unoccupied states and negative values to occupied states close to the Fermi
level.  Thus, the function $w^N(\text{\bf r})$ gives a {\em spatially resolved
picture} of the essential information contained in the local density of states
around the Fermi energy, which also forms the starting point for
tight-binding-like models of the local
reactivity~\cite{hof88,coh94a,ham95}. Moreover, the concept underlying the
definition of reactivity in Eq.~(2) is flexible; different reactivity indices
can be defined incorporating weight functions other than the Fermi function
used here for the occupation of Kohn-Sham eigenfunctions.

The determination of $w^N(\text{\bf r})$ requires two self-consistent
calculations of the electron density for different electron temperatures
$T_{\text{el}}$ {\em without} changing the number of electrons in the system,
a task readily performed. In contrast, the change of the electron temperature
at constant $\mu$ and {\em constant} external potential used defining
$w^\mu(\text{\bf r})$ results in a change of the number of electrons per unit
cell. The resulting charged systems cannot be treated within conventional
supercell geometries.  The finite difference formulation of equ.~(3) has in
addition the advantage that it enables the modelling of an increasing strength
of the interaction of the adsorbate with the surface by increasing the value
of $T_{\rm el}$.

As an example of the applicability of $w^N(\text{\bf r})$ to the study of the
reactivity of metal surfaces we consider the interaction of a hydrogen
molecule with the (100) surface of Pd.  The calculations presented below were
performed using DFT together with the generalized-gradient approximation
(GGA)~\cite{per92} and employing the full-potential linear-augmented plane
wave method~\cite{wien93,fhiv}.  The parameters used in the calculation
correspond to those used in Ref.~\onlinecite{wil99}.  The metal substrate is
modeled by five layers separated by a 10~{\AA} thick vacuum region.

The function $w^N(\text{\bf r})$ is displayed in Fig. 1 for clean Pd\,(100),
calculated using $k T_{\text{el}}$ =70 meV
and equ.~(\ref{wdef2}). 
The redistribution of electrons is quite pronounced.  There is a clearly
identifiable electron rearrangement within the $d$-shell: The $(x^2-y^2)$
electron density is increased (see Fig. 1 a,b) and the $(3z^2-r^2)$- (see
Fig. 1 c,d) and $(xy)$-orbitals (see Fig. 1 a) are depleted.  The
$(x^2-y^2)$-orbitals contribute to the bonding in the surface between the Pd
atoms, which can be inferred from the fact that their band is rather wide and
contains the highest number of unoccupied $d$ states at the surface.

\begin{figure}
\begin{picture}(-300,200)(100,250) 
\put(0,-50) { \includegraphics{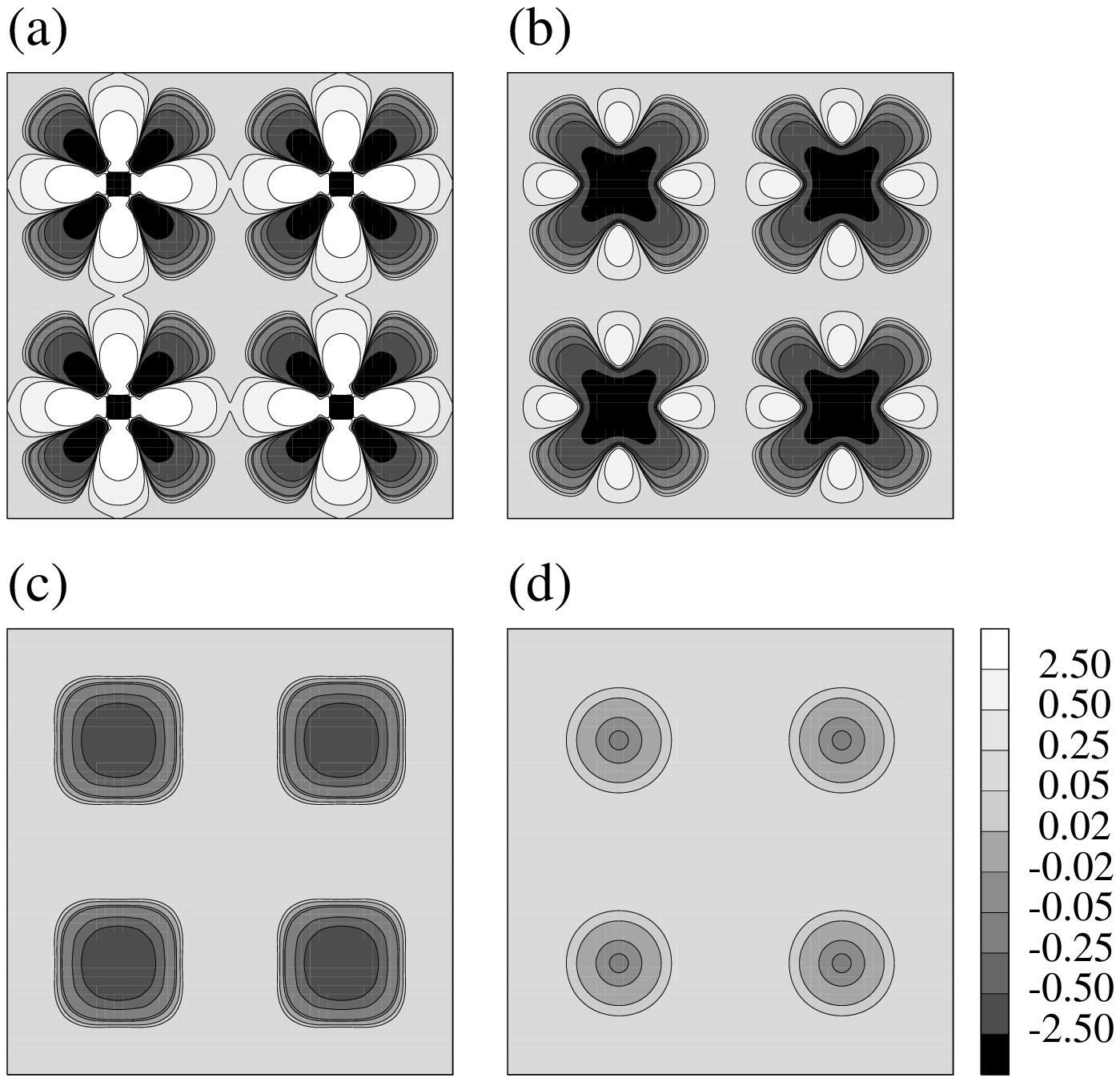}} 
\end{picture}
\caption{Contour plot of the function $w^N(\text{\bf r})$ for clean
Pd\,(100) along planes parallel to the surface at different heights $Z$ above
the center of the surface Pd atoms:
a) $Z$=0 {\AA}, b)  $Z$=0.5 {\AA}, c)  $Z$=1 {\AA}, and d)  $Z$=1.5 {\AA}.
Units are 10$^{-1}${\AA}$^{-3}$ eV$^{-2}$.}
\label{wr}
\end{figure}

We now consider the dissociative adsorption of a hydrogen molecule on the
Pd\,(100) surface in order to relate it to the reactivity index of Fig. 1.
The DFT calculations~\cite{wil99} show that a H$_2$ molecule at a constant
height, say at 1.8 \AA\, above the center of the top Pd layer, experiences a
significant corrugation of the potential energy as a function of its lateral
position (when the H$_2$ center of mass is at 1.8 \AA\, the hydrogen wave
functions extend up to $Z \approx 1.0$ \AA): The potential energy for the
H$_2$ center of mass at the hollow site is close to zero, at the bridge site
it is attractive by 0.12 eV, and at the on top site the attraction is 0.28 eV
(see Ref. \cite{wil99}).  This corrugation can be inferred, at least
qualitatively, from the $w^{N}$ contour plot of Fig. 1: The strongest decrease
of the surface electron density at distances between 1.0 and 1.5 \AA\, occurs
at the on top site.  Why does this strongest decrease of $w^{N}({\bf r})$
parallel the highest attraction for the molecule ?  The dissociative
adsorption of H$_2$ involves the breaking of the H-H bond and at the same time
the formation of new hydrogen-surface bonds~\cite{ham95,wil99}.  In a
bond-orbital picture~\cite{ham95} hydrogen dissociation is non-activated if
the H$_{2}$ $\sigma_g$-orbital as well as the $\sigma_{u}$ interact with the
occupied $d$-states, forming bonding states which are occupied (below the $d$
band), and if the corresponding antibonding orbitals remain essentially empty,
{\em i.e.}, if their DOSs lie essentially above the Fermi level.  In
particular the partial occupation of the $\sigma_{u}$ state destabilizes the
H-H bond.  It is quite plausible that this scenario is most effective at a
site of the metal surface with a high density of $d$ states just below the
Fermi level, which means at regions with large negative values of
$w^N(\text{\bf r})$.  Indeed, the spatial distribution of $w^N(\text{\bf r})$
in Fig.~\ref{wr}d suggests that this region is at the on-top position. Here
the H$_2$ molecule interacts strongly with the $(3z^2-r^2)$-orbital of the
surface Pd atom which gets partially depleted and electrons are transferred
into the $(x^2-y^2)$-orbitals located in the surface plane.

\begin{figure}
\psfig{figure=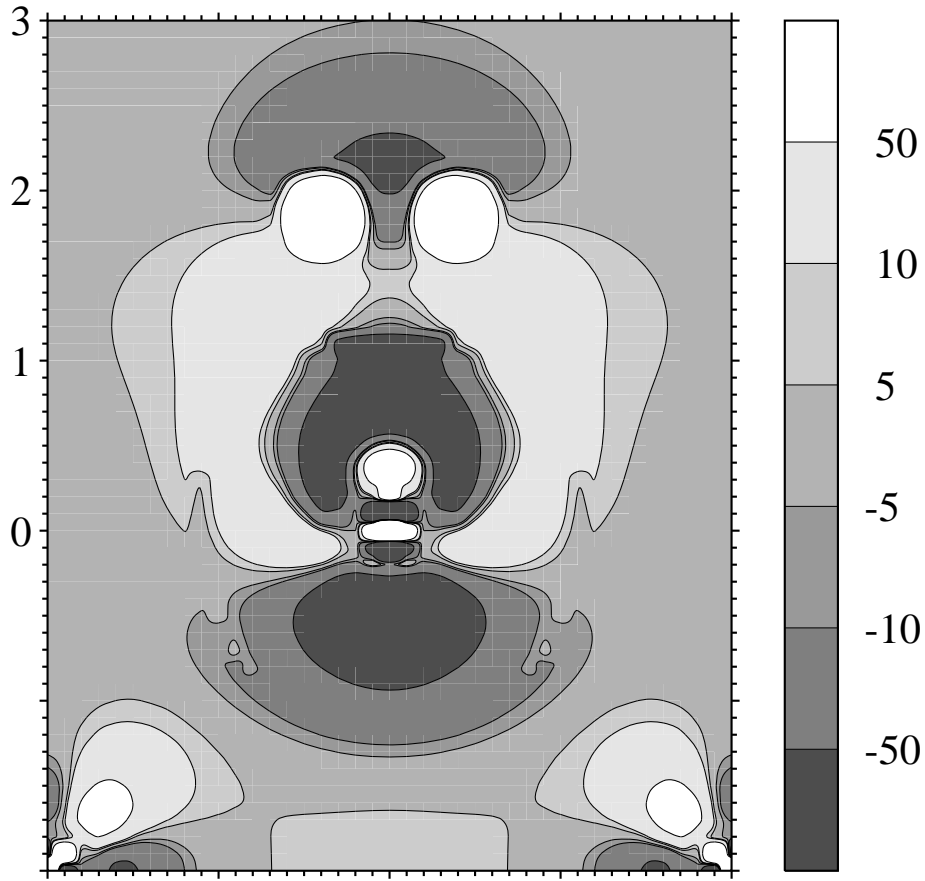}
\caption{Electron density change of a H$_2$ molecule above Pd\,(100) [see
text] along a the (011) plane. The molecule is placed at a height $Z= 1.8$
{\AA} above a surface Pd atom with the molecular axis parallel to the surface,
and the two H atoms point toward hollow sites.  Units are
10$^{-3}$/{\AA}$^3$.}
\label{h2top}
\end{figure}

Figure~\ref{h2top} displays DFT-GGA results for the interacting
H$_2$-Pd\,(100) system; shown is the electron density {\em change} for a
geometry with the hydrogen molecule at $Z=$ 1.8 {\AA} over the on top
position.  This {\em change} is defined as the electron density of the
H$_2$-Pd\,(100) system minus the density of the clean Pd\,(100) surface and
minus the density of a H$_2$ molecules with the same bond length.  It confirms
the just-mentioned character of the interaction, in particular the decrease of
the occupation of the ($3z^2-r^2$)-orbital of the surface Pd atom is clearly
visible.  All features in Fig.~\ref{h2top} are consistent with the spatial
variation of $w^N(\text{\bf r})$ and with the interpretation of hydrogen
dissociation in terms of a tight-binding picture~\cite{ham95}.

\begin{figure}
\psfig{figure=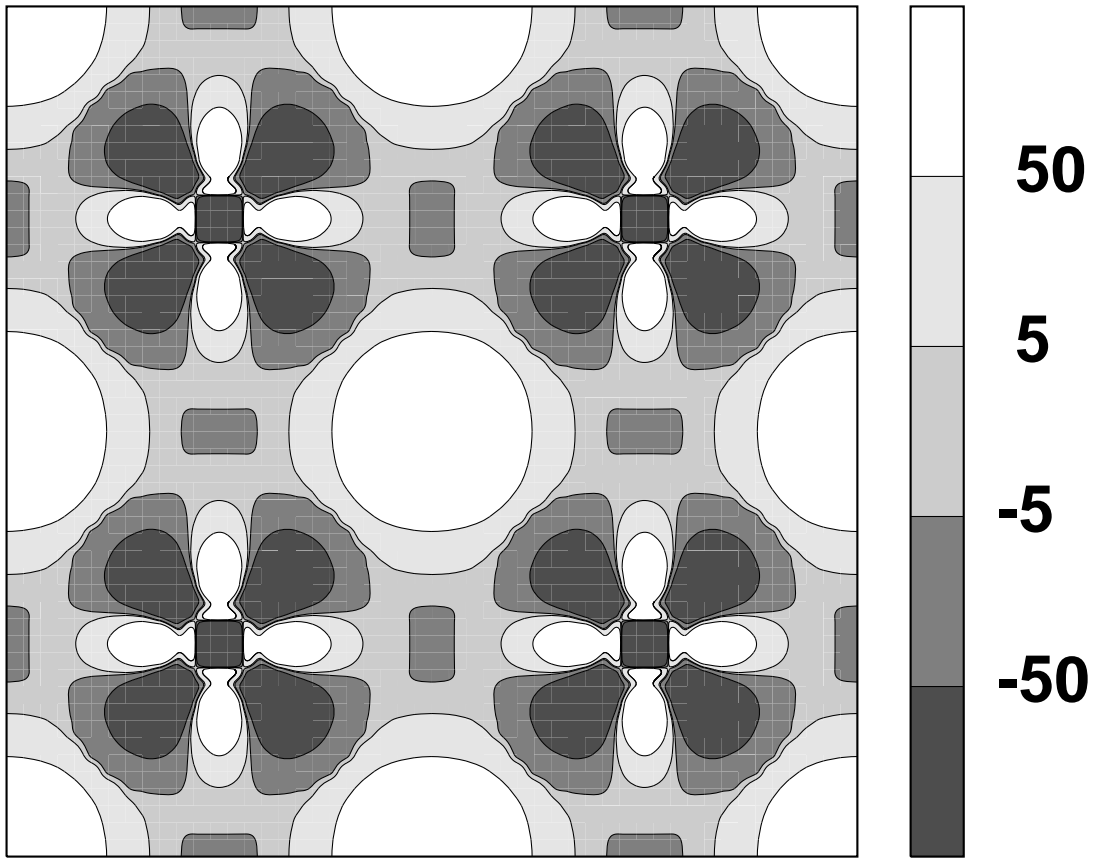}
\caption{Electron-density change at Pd\,(100) due to the
adsorption of one monolayer of hydrogen at the surface hollow sites.
The plane is the same as in Fig.~1a,
the units are 10$^{-3}$/{\AA}$^3$.}  
\label{hol}
\end{figure}

While the reactivity concept is based on perturbation theory~\cite{coh96a} and
can thus give quantitative information only for weakly interacting systems, it
is not unreasonable to expect that for strong interactions the concept may
still give {\em qualitatively} correct guidelines.  Indeed, we find that for
the chemisorption of atomic hydrogen at Pd\,(100) the function $w^{N}$ can
explain why the favorable adsorbate site is the fourfold hollow.  The DFT
calculations~\cite{wil94} of this chemisorption system tell that the hollow
site adsorption is by about 0.3 eV more favorable than the bridge and 0.6 eV
more favorable than the on-top site adsorption.  Following the same lines as
used for the discussion for the H$_2$ molecule we expect that the H atom
should prefer a site at which it has good overlap with dark (i.e. negative)
sections of $w^{N}({\bf r})$. Because of the smallness of the H-atom, this is
the hollow site (compare Fig. 1), in good agreement with the results of the
DFT calculations of the interaction system.  Figure~\ref{hol} gives the
calculated electron density change induced by a monolayer of H adatoms placed
in the surface hollow sites. It compares surprisingly well with the prediction
of the local polarization in Fig.~\ref{wr}a.
The comparison  shows that the spatial
variation of the function $w^N(\text{\bf r})$ is directly correlated with the
shape of the electron polarization introduced by the adsorbed
hydrogen.

In conclusion, the function $w^{N}({\bf r})$ [see equ. 2, 3] of the clean
surface is introduced as an isoelectronic reactivity index supplementing the
local softness. It is closely related to the local polarizability of valence
electrons induced by an atom or molecule chemically interacting with the
surface.  For the example of the dissociative adsorption of H$_2$ molecules on
Pd\,(100), we show that the spatial variation of the function $w^N(\text{\bf
r})$ allows rationalization of the preferred dissociation pathways as
well as the preferred chemisorption sites of H adatoms. The above discussion
was done for metal substrates.  However, the same reasoning applies also for
semiconductors and insulators.

\end{document}